\newif\ifdraft
\drafttrue

\newif\ifeasychair
\easychairtrue

\newif\iffinal
\finaltrue

\ifdraft
\documentclass[draft]{llncs}
\else
\documentclass{llncs}
\fi

\usepackage{savesym}

\usepackage[T1]{fontenc}
\usepackage[utf8]{inputenc}
\usepackage{lmodern}
\usepackage[scaled=.8]{beramono}
\usepackage{latexsym}

\usepackage[final]{graphicx}
\usepackage{amssymb}
\usepackage{mathtools}
\usepackage{bm}
\usepackage{comment}
\usepackage{currfile} 
\usepackage{listings}
\usepackage{paralist} 
\usepackage[inline]{enumitem} 

\usepackage{tikz}
\usetikzlibrary{calc,positioning,arrows,fit}

\savesymbol{todo}
\usepackage[show]{ed}
\restoresymbol{ed}{todo}
\usepackage[obeyDraft,textsize=tiny]{todonotes} 

\usepackage{verbatim}

\usepackage[british]{babel}
\usepackage[babel]{csquotes}
\MakeAutoQuote{“}{”}
\MakeAutoQuote*{‘}{’}

\usepackage[backend=bibtex,hyperref=auto,
firstinits=true,style=numeric,
urldate=iso8601,isbn=false,doi=false]{biblatex}
\ifeasychair
\else
\fi
\DeclareNameAlias{author}{last-first}
\DeclareNameAlias{editor}{last-first}
\DeclareNameAlias{translator}{last-first}
\DeclareFieldFormat{labelnumberwidth}{#1.}
\DeclareFieldFormat{title}{#1}
\DeclareFieldFormat
  [article,inbook,incollection,inproceedings,patent,thesis,unpublished]
  {title}{#1}
\DeclareFieldFormat{journaltitle}{#1}
\newbibmacro*{volume+number+eid}{%
  \printfield{volume}%
  \iffieldundef{number}
    {}
    {(%
      \printfield{number}%
     )%
    }%
  \setunit{\addcomma\space}%
  \printfield{eid}}
\DeclareFieldFormat{url}{\url{#1}}

\usepackage{url}

\newcommand{\href}[2]{#2}

\DeclareMathOperator*{\argmax}{arg\,max}

\begin{document}

\title{Proving soundness of combinatorial Vickrey auctions and generating verified executable code\iffinal\thanks{
This work has been supported by EPSRC grant EP/J007498/1 and an LMS Computer Science Small Grant.  We are grateful to Peter Cramton, Elizabeth Baldwin, and Paul Klemperer for their insights and guidance; Rowat thanks Birkbeck for its hospitality.}\fi}

\author{Marco B.\ Caminati\inst{1}
\and Manfred Kerber\inst{2}
\and Christoph Lange\inst{2}
\and Colin Rowat\inst{3}
\institute{%
\texttt{http://caminati.net.tf}, Italy
\and Computer Science, University of Birmingham, UK
\and Economics, University of Birmingham, UK
\texttt{http://www.cs.bham.ac.uk/research/projects/formare/code/auction-theory/}
}}

\maketitle

\begin{abstract}
  Using mechanised reasoning we prove that combinatorial Vickrey auctions are soundly specified in that they associate a unique outcome (allocation and transfers) to any valid input (bids).  Having done so, we auto-generate verified executable code from the formally defined auction.  This removes a source of error in implementing the auction design.  We intend to use formal methods to verify new auction designs.  Here, our contribution is to introduce and demonstrate the use of formal methods for auction verification in the familiar setting of a well-known auction.
\end{abstract}


\section{Introduction} \label{sec:introduction}

This paper presents a unified approach to two important questions in
auction design and practice: \begin{enumerate*}\item the question of sound specification (\emph{for any valid set of inputs, i.e.\ bids, does the auction result in a unique outcome, i.e.\ allocation and transfers?}), and \item that of reliable implementation of the specification (\emph{is the auction software a faithful implementation of the auction design?}).\end{enumerate*}  Failure on either front can be hugely costly -- resulting in both litigation and reputational damage -- especially in high-stakes, one-off auctions.  %

Therefore, novel auction designs are typically extensively tested before implementation.  In simple cases, theoretical results may exist (e.g.\ the well-known revenue equivalence results, or Vickrey's theorem about truth-telling as a weakly dominant strategy).  Further, there are no well-known errors in any manually-derived theorems in auction theory, nor do leading auction theorists doubt important results' soundness.  Rather, their greater concern -- especially for more complicated auctions, or auctions running under less restrictive conditions than theory typically assumes (e.g.\ risk-neutrality, or common knowledge of the support of bidders' valuations) -- is that theory is insufficient.  In these cases, auction designs and their software implementations are tested on test data (as, e.g., generated by CATS~\cite{le-br-sh-06}), in experimental labs, and as “dry runs” with the intended bidders.  Such tests may add considerably to confidence in an auction, but merely demonstrate that no anomalies have been discovered yet on the finite set of test cases to which the auction has been exposed.%
\footnote{Dijkstra famously said that “testing shows the presence, not the absence of bugs.”~\cite
{buxton1970software}}

When more confidence is desired, it may be possible to prove 
certain properties using mechanised methods.  In the same way that computer algebra software such as Mathematica 
confirms algebraic calculations or solves algebraic problems, proof assistants can be used to confirm logical operations, or even to generate proofs.

Early hopes that mechanised reasoning would solve open mathematical problems proved too optimistic, with only three important results falling to mechanised means.  First, over 100 years after the four-colour map theorem was posed (four colours suffice to colour any planar map without any adjacent regions sharing a colour),
it was proved by a combination of mathematical text and
a purpose-built 
computer program%
~\cite{gon-08}.
Despite concerns about relying on such “black box” code, the proof has been grudgingly accepted, and has subsequently been confirmed by Coq, a general-purpose theorem prover%
~\cite{gon-08}.  Second, Robbins' 
conjecture that one of the axioms in Huntington's basis for Boolean algebras was equivalent to one of his axioms was resolved after 60 years when a solver generated -- after running for eight days -- a 17-step proof that humans could check manually; fine-tuning found an eight-step proof in five days~\cite{mcc-97}.  Third, Kepler's conjecture about optimal sphere packing was reduced, by the mid-20th century, to exploring about $5{,}000$ possible packings.  Hales solved that by minimising a 150-variable function on each packing, generating some $100{,}000$ linear programming problems~\cite{hal-05}.  As 12 human referees could only -- after a five-year review -- report that they were “99\% certain” the proof was correct, Hales set out to use the HOL Light general-purpose theorem prover in the same way that Coq had been used to prove the four-colour map theorem.

Outside of pure mathematics, theorem provers have been more widespread.  Following an embarrassing and costly recall in the mid-1990s, Intel has used theorem provers to confirm that, for example, its chips satisfy the IEEE floating point division standard: viewing a 
chip as a set of Boolean statements, a prover may ask whether particular theorems hold within the world defined by those statements~\cite{harrison-sfm}.  Similarly, any computer program defines a logical universe in which theorem provers may assess the truth 
of certain statements: in code controlling automated 
commuter rail systems, e.g., the theorem that no two trains occupy the same location at the same time has been proven~\cite{wo-la-bi-fi-09}.  With Facebook's 2013 acquisition of Monoidics, a theorem-proving start-up firm, these techniques may be gaining greater public attention.

While the results are only as reliable as the proof assistant used, there is reason to believe that mechanised, or “formal”, methods can achieve higher levels of reliability than purely manual methods: as proof checkers are typically tested on a wide range of problems, their results are not subject to preconceptions about what the correct result should be in a particular problem.  Further, computers never tire of checking fine details, nor does their literalism allow them to implicitly infer hidden assumptions.

Section \ref{sec:our-approach} outlines how we addressed the two questions stated initially by formal methods; concretely, we verify combinatorial Vickrey auctions.  Section \ref{sec:isabelle} justifies our choice of the Isabelle/HOL prover.  Section \ref{se:VCG} provides a “paper” definition of Vickrey's auction.  Section \ref{se:s-spec} describes how we translated that into the prover's formal language and proved that the auction is soundly specified.  Section \ref{sec:gener-code-integr} describes the generation of verified code from the formalisation.  Section \ref{sec:related-work} discusses related work%
, and section \ref{se:concle} concludes.

\section{Our Approach}
\label{sec:our-approach}

We use the Isabelle/HOL\footnote{In the following, we use “Isabelle/HOL” only for referring to the specific choice of higher order logic (HOL) within the Isabelle theorem proving system.} prover~\cite{isabelle} to verify that the combinatorial Vickrey-Clarke-Groves (VCG) mechanism is soundly specified, and to automatically generate executable code that faithfully implements the formally specified design.\footnote{In the following, we refer to both the combinatorial VCG mechanism and its single good counterpart simply as “Vickrey's auction”.}  This auction is, of course, well understood (q.v.~\cite{au-mi-06}): there is no doubt that it is well defined, nor that it has been successfully implemented.  We work with Vickrey's auction for precisely this reason: introducing the novel techniques of mechanised reasoning and code generation in a simple and familiar environment.

\begin{figure}[ht]
\centering
\begin{tikzpicture}
  \node[draw,minimum height=7ex,align=center] (theorems) {\textbf{\ref{it:prop}.~Theorems}};

\node[below=1cm of theorems,draw,minimum height=7ex,align=center] (def) {\textbf{\ref{it:def}.~Definitions}};

  \node[above=0cm of theorems,align=center] (spec-label) {
    \scriptsize formal specification\\
    \scriptsize (written by Isabelle user, needs\\
    \scriptsize review by auction designer)};

  \node[fit=(spec-label) (theorems) (def),draw,dashed] (spec) {};

  \node[right=3cm of def,draw,minimum height=7ex,align=center] (code) {%
    \textbf{Code}\\
    \scriptsize (executable Scala)};

  \node[right=3cm of theorems,draw,minimum height=7ex,align=center] (proof) {%
    \textbf{\ref{it:proof}.~Proof}\\
    \scriptsize (\ref{it:check}.~checked by Isabelle)};

  \draw[->] (theorems) to node[left,align=right] {%
    \scriptsize state soundness\\
    \scriptsize and other properties of} (def);

  \draw[->] (code) to node[near start,above right,align=left] {%
    \scriptsize known to implement (by proof\\
    \scriptsize and by trusting code generator)} (spec);

  \draw[->] (def) to node[below,align=center] {%
    \scriptsize \ref{it:code}.~code generation\\
    \scriptsize (Isabelle)} (code);

  \draw[->] (proof) to node[below,align=center] {\scriptsize proves} (theorems);
\end{tikzpicture}
\caption{High-level outline of our approach}
\label{fig:high-level-approach}
\end{figure}
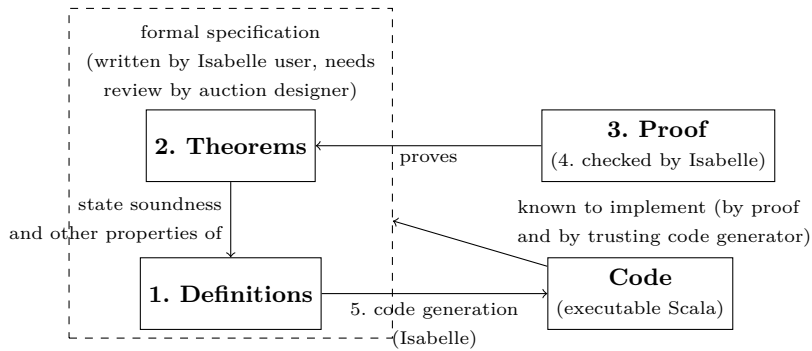

We proceed in five steps.  \begin{enumerate*}\item\label{it:def} We \emph{define} a Vickrey auction 
and \item\label{it:prop} \emph{specify} the desired properties of a Vickrey auction, both in Isabelle/HOL's formal input language.  Both steps require human expertise to ensure that the formal definitions and specifications correspond to the auction designer's intuition.  Once this is done, \item\label{it:proof} \emph{proofs} that the auction possesses its specified properties are manually written in Isabelle/HOL's language.\footnote{Technically, this is \emph{interactive} rather than \emph{automated theorem proving}, which could generate proofs autonomously.   Section~\ref{sec:isabelle} further comments on this difference.}  Here, the property of interest is that the Vickrey auction -- viewed as a relation from inputs (bids) to outcomes (allocations and transfers) -- is a function, associating a unique outcome to every valid input.

\item\label{it:check} Isabelle 
then \emph{formally checks} the proofs, producing a \emph{proof object} (also called “certificate”) to confirm that the manual proof is correct, or to expose faulty steps in the proof.  \item\label{it:code} If the Isabelle/HOL definitions have been written in a computable way rather than implicitly, they define the auction as an \emph{algorithm}, which Isabelle's \emph{code generator} can automatically export to a variety of executable languages, including the Java-based Scala.  Isabelle/HOL supports both algorithmic and implicit definitions; e.g., one can state and prove the existence of the maximum element of some set without explicitly computing its value.
Implicitly defining a concept, such as the winners of an auction, without supplying an algorithm, e.g.\ to solve the winner determination problem (WDP), is usually more straightforward for an Isabelle user, the resulting formalisation is close to textbook style and thus easier to review for auction designers, and properties of concepts defined that way can be proved more easily in Isabelle using classical reasoning.\footnote{Proofs in classical logic, in contrast to constructive proofs, allow the law of the excluded middle and thus proof by contradiction and therefore are usually shorter.}  Generating verified executable code, however, does require algorithmic definitions.  We therefore develop two parallel formalisations: using the previous example, we specify the concept of the winners of an auction in the implicit style an auction designer would also use on paper (“the winners correspond to a value-maximising set of bids”), and we provide an algorithm for solving the WDP, and then prove that it satisfies the specification, i.e.\ that its result set is equivalent to the implicit definition.  To keep figure~\ref{fig:high-level-approach} simple, we detail this separately, with concrete examples, in figure~\ref{fig:paper-vs-exec}.  While this approach involves a duplication of effort, it has the big advantage that the correctness of the generated code depends exclusively on the definitions (specification) and the correctness of the Isabelle system itself.\footnote{The alternative is to use a purely constructive logic.  In the Coq and NuPRL theorem provers, for example, it would be possible to just provide the paper-like specification of the winning allocation as a definition, then, constructively, prove the soundness of that definition, and then extract the algorithmic content of that proof (which solves the WDP), i.e.\ executing the proof itself.  While avoiding our duplication of effort, the constructive approach not only has the disadvantage of making proofs more difficult, but also making  the development of verified efficient algorithms harder, since efficiency has to be achieved by developing correspondingly usable proofs.  This is less direct than writing down specification and code separately and then proving that the code meets the requirements of the specification.}
\end{enumerate*}

\section{The Isabelle/HOL theorem prover}\label{sec:isabelle}

We initially chose Isabelle/HOL because an earlier system comparison study based on the proof of Vickrey's theorem for single-good auctions proved its general suitability for auction design, taking into account criteria such as the expressiveness of its language, its combination of interactive and automated theorem proving, its library coverage, and the comprehensibility of its input and output language for auction designers~\cite{LangeEtAl:CompProvAuctThy13}.

Isabelle has historically been seen as an interactive, rather than automated, theorem prover, meaning that it checks proof steps written by human users rather than autonomously generating proofs.  The “interactive vs.\ automated” distinction has recently been blurring in 
theorem provers, with each adopting features of the other~\cite{LangeEtAl:CompProvAuctThy13}.  Isabelle has three automated tools that we use; in practice, they work best for small reasoning steps and time out otherwise.
\begin{enumerate*}
  \item Sledgehammer~\cite{isabelle-sledgehammer} is an interface to automated theorem provers installed locally or available as web services.
  \item Nitpick~\cite{isabelle-nitpick} generates counterexamples and outputs them in a human-readable notation.
  \item The \texttt{try} command~\cite[section~12.10]{isabelle-isar} provides an integrated frontend, applying Sledgehammer, Nitpick and other built-in automated provers and disprovers to a proof goal.
\end{enumerate*} %

 %

{
  \centering
  \ifeasychair
  \input{3-classical-logics.tkz}
  \else
  \input{../../lib/graphics/logic/3-classical-logics.tkz}
  \fi
}

One reason for the historical view of Isabelle as an interactive theorem prover is that it implements higher-order logic (HOL), the most expressive of the three 
common classical logics briefly illustrated above.  One common trait is \emph{soundness}: if a property, $\varphi$, is deduced from a system $\Gamma$ of statements by proof steps (written $\Gamma \vdash \varphi$), it is -- in fact -- a property of the system (written $\Gamma \models \varphi$).

Propositional logic is 
only able to express concrete, finite statements: given fixed bids, for example, 
one can express 
that one exceeds the other.  This 
simplicity entails
advantages%
: the logic is \emph{complete}, meaning that any question can, in principle, be answered by skillful use of the logic's calculus in proof steps.  Formally, $\Gamma \models \varphi$ implies $\Gamma \vdash \varphi$.  The logic is also \emph{decidable}: if a statement is either true or false, then there is an algorithm for deriving its truth.

First-order logic (FOL) is more expressive than propositional logic, adding the universal (for all, $\forall$) and existential (there exists, $\exists$) quantifiers, as well as predicates and functions, to the former's concrete objects.  
Gödel's completeness theorem proved that it is complete.  Thus it may be possible to use fully automated theorem proving (ATP) to derive answers to questions~\cite[31]{wo-la-bi-fi-09}.  Against this, though, FOL is no longer decidable:
while completeness makes sure a formal, finite proof exists for any true statement (this is called “semi-decidability”),
no universal algorithm exists to solve the preliminary problem of establishing the truthfulness of any given statement.

Finally, higher order logic (HOL) adds quantification over functions and predicates to FOL, making it extremely expressive.    Gödel's 
incompleteness theorem proved, though, that it is not even complete.  
As Isabelle implements HOL, it is theoretically capable of expressing questions that simply cannot be answered -- ruling out the general possibility of applying automation.  This hardly affects our practical work: we are not posing “strong” HOL statements, e.g., on the existence of functions that satisfy certain abstract properties, but are merely asking whether \emph{given} functions, which describe an auction, have certain concrete properties.  We prefer HOL as it allows for naturally expressing the essential concepts of auctions.  When modelling bids as functions from a bidder and a set of goods to a price, FOL would not support statements about all possible bids.

Not only is Isabelle/HOL's language expressive, but it looks close to mathematical textbook notation and therefore, after some training, allows anyone with a mathematical background to grasp the essentials of an Isabelle/HOL formalisation
;  section~\ref{se:s-spec} gives examples.  However, the level of detail and explicitness required for machine-checkable proofs, and for obtaining machine-generated counter-examples, is much higher than on paper~\cite{LangeEtAl:CompProvAuctThy13}.  Thus, writing such a proof in the first place is labour intensive and requires deep understanding of Isabelle.  Those proofs that Isabelle can generate automatically are even harder to read for non-experts.

Finally, Isabelle's ability to generate executable code in Scala was a key criterion.  Scala code runs on the Java Virtual Machine, can invoke Java code (and vice versa), and has access to the rich libraries available for Java, including web services, user interfaces and databases.  Generating verified Scala code therefore promises easy integration into business software; in fact, Scala is increasingly being used for business-critical software.%

\section{Combinatorial VCG auctions} \label{se:VCG}

Our description of a multi-good 
or combinatorial VCG auction follows~\cite{au-mi-06}: let the set of \emph{agents} be $\left\{ 0, \ldots, N \right\}$, with agent $0$ denoted as the \emph{seller} and the rest as \emph{bidders}.  The seller's \emph{endowment} is the set $X_0 \ne \emptyset$ of indivisible goods.  An \emph{allocation} is a partition of the endowment, $X_1, \ldots, X_N$, where component $X_n$ denotes bidder $n$'s share of the allocation.  An admissible allocation does not distribute more goods than the seller has available, but possibly fewer ones, so that $\bigcup_{n=1}^N X_n \subseteq X_0$.
Any given bidder $n$ privately values each subset of the endowment $X \subseteq X_0$ at $v_n \left( X \right)$\footnote{Our present focus is not on efficiently representing valuations (cf.\ \cite[part~II]{cr-sh-st-06}).}; we assume $v_n \left( \emptyset \right)=0$.  Bidders simultaneously submit \emph{bids} to the seller, $b_n \left( X \right), \forall X \subseteq X_0$.\footnote{In practice, the combinatorial task of supplying bids for all subsets of $X_0$ is a problem that dynamic auctions seek to mitigate; it does not play a role in the present analysis.}  In general, $b_n \left( X \right)$ may depend on a bidder's own valuation 
as well as $n$'s estimates of the others' values.\footnote{Our formalisation does not make this probabilistic relation between bids and values explicit but starts with the bids, skipping the values.}

We assume that the seller seeks to maximise value, as proxied by bids.  The seller therefore solves for the \emph{value-maximising allocation}, 

\begin{equation} \label{eq:wdp}
  X^\ast \in \argmax_{ X_1, \ldots, X_N} \sum_{n=1}^N b_n \left( X_n \right) \text{ s.t. } \bigcup_{n=1}^N X_n \subseteq X_0 \text{ and }  n \neq n' \text{ iff } X_n \cap X_{n'} = \emptyset
\end{equation}

at prices
\begin{equation} \label{eq:2price}
  p_n \equiv \alpha_n - \sum_{m \ne n} b_m \left( X^\ast_m \right)
\end{equation}
where
\[
  \alpha_n \equiv \max_{\substack{X_m\\ m=1,\ldots,N, m \ne n}}  \left\{ \sum_{m \ne n} b_m \left( X_m \right) \left| \bigcup_{m \ne n} X_m \subseteq X_0 \right. \right\}
\]
is the value reportedly generated by the value maximisation problem when solved without $n$'s bids.  Thus, $p_n$ is the opportunity cost of the items won by bidder $n$ -- their reported value to the remaining set of bidders, net of their value had $n$ not bid.
Equation \eqref{eq:wdp} is the \emph{winner determination problem} (WDP).  When there are multiple solutions $X^\ast$ (we denote the set of them by $\mathcal{X}^\ast$), a tie-breaking rule must be used to adjudicate.  A typical solution involves assigning random numbers to each possible bundle, denoted by $r_n \left( X \right)$, including the empty set, $r_n \left( \emptyset \right)$.  The WDP is then run once more over $\mathcal{X}^\ast$ to maximise the sum of the random numbers:
\begin{equation}
  Y \equiv \argmax_{X^\ast \in \mathcal{X}^\ast} \sum_{n=1}^N r_n \left( X_n \right).
\end{equation}

When the random numbers are drawn from the same distribution, the tie-breaking is neutral in the sense of summing over $n$ draws in all cases.

We conclude by fixing these concepts in the context of two concrete examples.

\begin{example}[{\cite[23]{au-mi-06}}]
  Assume $N = 3$ bidders submit the following bids for two items $A$ and $B$:
  \[
    b_1 \left( AB \right) = b_2 \left( A \right) = b_2 \left( B \right) = b_3 \left( A \right) = b_3 \left( B \right) = 2;
  \]
  and $b_n \left( X \right) = 0$ for all other $n \in \left\{ 1, 2, 3 \right\}$ and item sets $X \subseteq \left\{ A, B \right\}$.  This leads to two equivalent value-maximising allocations; we focus on $X^\ast = \left( \emptyset, A, B \right)$, omitting the seller.\footnote{We use $\left( \emptyset, A, B \right)$ as a shorthand notation for $\bm{x}^\ast = \left\{ \emptyset, \{A\}, \{B\} \right\}$.}  Then
  \begin{align*}
    \alpha_1 & = \max_{X_2, X_3} \left\{ b_2 \left( X_2 \right) + b_3 \left( X_3 \right) \left| X_2 + X_3 \subseteq X_0 \right. \right\} = b_2 \left( A \right) + b_3 \left( B \right) = 2 + 2 = 4; \text{ and}\\
    \alpha_2 & = \max_{X_1, X_3}  \left\{ b_1 \left( X_1 \right) + b_3 \left( X_3 \right) \left| X_1 + X_3 \subseteq X_0 \right. \right\} = 2;
  \end{align*}
  where $\alpha_2$ may be determined by either $b_1 \left( 0 \right) + b_3 \left( B \right) = 2$ or $b_1 \left( AB \right) + b_3 \left( 0 \right) = 2$.  Finally, by symmetry, $\alpha_3 = \alpha_2 = 2$.  Prices are then
  \begin{align*}
    p_1 & = 4 - \left[ b_2 \left( X^\ast_2 \right) + b_3 \left( X^\ast_3 \right) \right] = 4 - \left[ 2 + 2 \right] = 0; \\
    p_2 & = 2 - \left[ b_1 \left( X^\ast_1 \right) + b_3 \left( X^\ast_3 \right) \right] = 2 - \left[ 0 + 2 \right] = 0;
  \end{align*}
  and, by symmetry, $p_3 = p_2 = 0$.
\end{example}

In the special case of a single good, it is well known that the combinatorial Vickrey's auction awards the good to the highest bidder, who pays a price equal to the highest remaining bid; no other bidder pays anything.

\begin{example}[A single good as a special case of the $n$-good case]
  Assume
  \begin{itemize}
  \item a single indivisible good $A$, i.e.\ $X_0 = A$, and
  \item $N$ bidders, each with bids $b_n \left( A \right) = b_n$, and $b_n \left( \emptyset \right) = 0$.
  \end{itemize}
  Let $i$ be one of the bidders with the highest bids, i.e.\ $b_i=\max_{n=1}^Nb_n$.
  Then, $X^\ast_i = A$ and $X^\ast_j = \emptyset$ for all $j\ne i$, and $\alpha_k=\max_{j \ne k} b_j$ for each bidder $k=1,\dots,N$.\footnote{Any of the sums in the definition of $\alpha$ only consists of at most one non-negative summand: the allocation that awards the good to some bidder $j$ ($x_j=1$), and for which $j$ bid $b_j$.}  
  Prices are $p_k=\alpha_k-\sum_{j \ne k} b_j \left( X^\ast_j \right)$; concretely,
$
    p_i 
=\max_{j \ne i} b_j \text{ and}
$
$
    p_j 
=b_i - b_i = 0 \text{ for all }j \ne i\text{.}
$
\end{example}

\section{Proving that Vickrey's auction is soundly specified} \label{se:s-spec}

With the paper definitions in place, we now turn to demonstrating that Vickrey's auction is soundly specified, associating a unique, well-defined outcome (an allocation and transfers) to every valid input (bids).  More formally, when viewed as a relation from inputs to outcomes, Vickrey's auction is left-total and right-unique, i.e.\ a function.   We prove three intermediate lemmas, whose conjunction yields the result that Vickrey's auction is a well-defined function: 

\begin{enumerate}
  \item \label{RefTotality} “For each admissible input, the auction yields an outcome.”
  \item \label{RefCompliance} “Any outcome of the auction -- if one exists -- is well-defined”, meaning that
  the allocation vector does not allocate the same good more than once, and does not exceed the set of available goods;
  and that each bidder's payment is non-negative.
  
  \item \label{RefUniq} “Any outcome of the auction -- if one exists -- is unique.”
\end{enumerate}

We decompose our ultimate goal into these three intermediate goals for two reasons.  First, each of the three statements has a different structure %
\footnote{For $R(i,o)$ being the auction relation of an input $i$ to an outcome $o$, the formal structures of the statements are
  \begin{enumerate*}
  \item $\forall i . \mathit{adm}(i) \rightarrow \exists o . R(i,o)$,
  \item $\forall i \forall o . R(i,o) \rightarrow \mathit{wd}(o)$, and
  \item $\forall i \forall o_1 \forall o_2 . R(i,o_1) \wedge R(i,o_2) \rightarrow o_1=o_2$%
  \end{enumerate*}.
} and therefore requires a differently structured proof.  Second, we seek to not just accomplish a one-off soundness proof of one auction, but to develop a toolbox of  modular, reusable formalisations about auctions (cf.\ section~\ref{sec:gener-code-integr}): as proofs of further properties of an auction may rely on some aspects of the overall soundness proof but not on others, that toolbox should be modular.\footnote{For example, in our earlier realisation of this approach for the single-good Vickrey auction, the proof of Vickrey's theorem makes use of the – now proved – fact that the auction has a well-defined outcome.}

It turns out that for the combinatorial Vickrey auction goal \ref{RefTotality} and \ref{RefUniq} can be established in a straightforward way, as the Isabelle/HOL language allows concepts to be functions by definition.\footnote{As Isabelle/HOL is a typed language, this requires some discipline w.r.t.\ making sure that functions are applied to arguments of admissible types and return a value of the intended type.}  Equation \eqref{eq:2price} defines the transfers as a function of the winning allocation; as long as the tie-breaking rule is definable as a function, it picks a unique allocation; finally, goal \ref{RefTotality} requires us to show that a value-maximising allocation exists at all.  From the implicit, textbook-style definition ($X^\ast\in\argmax\sum\dots$) this is not obvious at a glance; however, as we explained in step \ref{it:code} of section~\ref{sec:our-approach}, the code generation objective requires us to define an algorithm that computes the set of value-maximising allocations.  Any such algorithm reasonably assumes that a certain starting allocation (which always exists, given non-empty sets of bidders and goods) attains the maximum value, and would then search for a better allocation, but in any case it is definable as an Isabelle/HOL function that returns a set of at least one allocation.   It remains to be shown that this function computes the same allocations as those in the set $\argmax\sum\dots$.
Goal~\ref{RefCompliance} is also straightforward to prove: our definition of value-maximising allocations precludes over-allocation by construction.

We state and prove soundness for a formalisation that is close to the original textbook style of the definitions in section~\ref{se:VCG}.  We formalised the essential concepts in three layers, from foundational mathematical structures to high-level concepts specific to auction theory: partitions of a set (of goods to be auctioned), allocations of sets of goods (i.e.\ members of one such partition) to bidders, and finally value-maximising allocations, i.e.\ solutions to the WDP.  This is the paper-like Isabelle definition of the set of possible allocations as the set of all relations between partitions of the set of goods and bidders that are injective functions:

\[
  \begin{array}{l}
    \mathbf{definition}\ \mathit{possible\_allocations} :: \text{``} \mathit{goods} \Rightarrow \mathit{bidder\ set} \Rightarrow \mathit{allocation\ set} \text{''} \\
    \mathbf{where}\ \text{``} \mathit{possible\_allocations}\ G\ N = \{ \mathit{potential\_buyer}\ .\ \exists P \in \mathit{partitions}\ G\ .\\
    \quad\wedge \mathit{Domain}\ \mathit{potential\_buyer} \subseteq P \wedge \mathit{Range}\ \mathit{potential\_buyer} \subseteq N\\
    \quad\wedge \mathit{right\_unique}\ \mathit{potential\_buyer} \wedge \mathit{injective}\ \mathit{potential\_buyer} \} \text{''}
  \end{array}
\]

\section{Generating verified code} \label{sec:gener-code-integr}

As explained in section~\ref{sec:our-approach}, Isabelle cannot generate code from implicit definitions such as the one stated above; instead, we had to provide a separate, algorithmic definition of each relevant concept in functional programming style:

\[
  \begin{array}{l}
    \mathbf{fun}\ \mathit{possible\_allocations\_alg} :: \text{``} \mathit{goods} \Rightarrow \mathit{bidder\ set} \Rightarrow \mathit{allocation\ list} \text{''} \\
    \mathbf{where}\ \text{``} \mathit{possible\_allocations\_alg}\ G\ N = \mathit{concat}\ [ \\
    \qquad [\ \mathit{potential\_buyer}\ .\\
    \quad\qquad\mathit{potential\_buyer} \leftarrow \mathit{injective\_functions\_list}\ Y\ (\mathit{sorted\_list\_of\_set}\ N)\ ]\\
    \quad.\ Y \leftarrow \textit{partitions\_alg}\ (\textit{sorted\_list\_of\_set}\ G)\ ]\text{''}
  \end{array}
\]

Instead of pulling into existence arbitrary partitions of the set of goods and arbitrary injective functions from such partitions to bidders, and then keeping those that meet the desired criteria, we invoke further algorithms computing the exact lists of desired objects.  Lists are often easier to compute with than sets; in particular, sets have no order and therefore do not allow for defining algorithms that pick one element\footnote{As we are dealing with finite sets, this is not a question of believing in the axiom of choice.  Of course we can pick \emph{some} element out of a finite set, but only an order on the set – easiest to obtain by converting it into a sorted list – guarantees that repeated executions of an algorithm always pick the \emph{same} element.} and then recursively process the remaining ones.

Next, we need to prove the equivalence of both definitions, concretely, that the list computed by the algorithm contains exactly the same allocations as the set defined implicitly.  In other words, we need to ensure that the algorithm implements its (complete) specification.  The proof, which is currently work in progress, works by induction over the numbers of goods and bidders.

\begin{figure}
  \centering
  \ifeasychair
  \includegraphics[width=\textwidth]{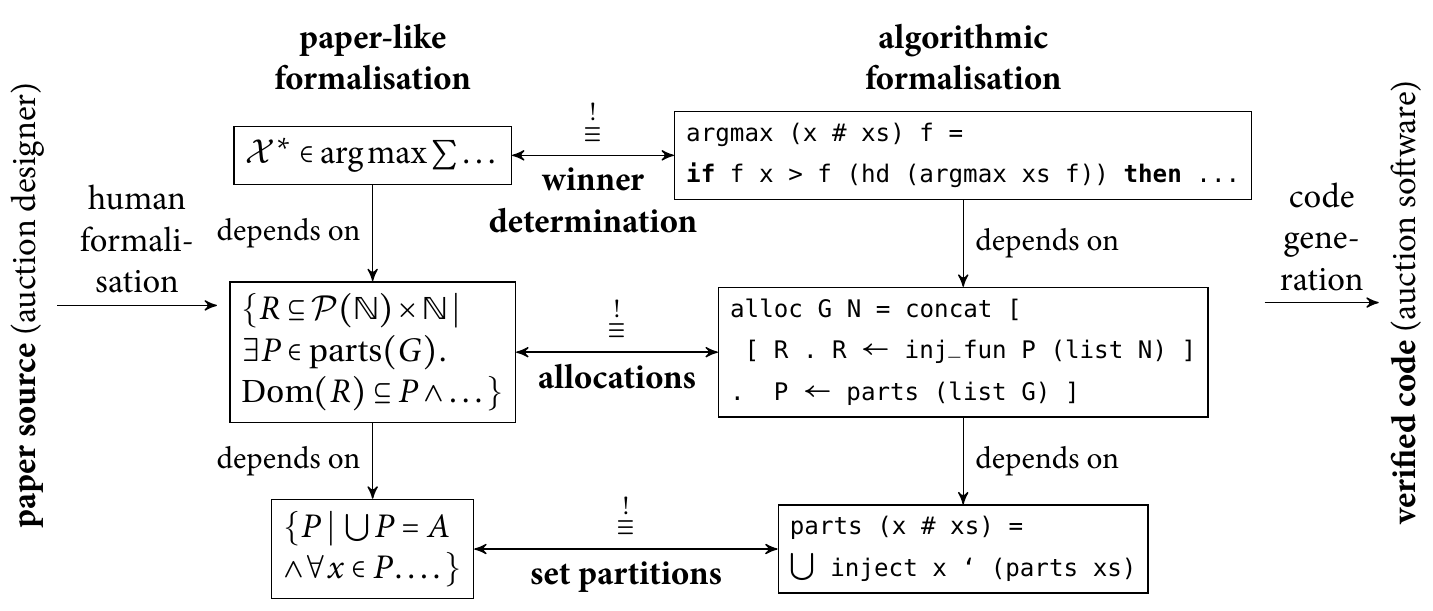}
  \else
  \includegraphics[width=\textwidth]{../../lib/graphics/auctions/wd-check}
  \fi
  \caption{Paper-like vs.\ algorithmic formalisation (definitions abridged; $\stackrel{!}{\equiv}$ denotes an obligation to prove equivalence)}
  \label{fig:paper-vs-exec}
\end{figure}
On the two other formalisation layers there are analogous pairs of paper-like and algorithmic formalisations proved to be equivalent, as shown in figure~\ref{fig:paper-vs-exec}.  If the paper-like formalisation corresponds to the auction designer's intuition and if we trust the code generator, we know, by these equivalence proofs, that the generated code faithfully implements the auction.  If we additionally know that the auction is sound (via theorems and proofs about the concepts defined, usually in the paper-like formalisation, as it easiest to do proofs there), we know that the generated code is correct, as explained in figure~\ref{fig:high-level-approach}.  In fact, the generated code possesses \emph{any} property that we prove of the paper-like formalisation – think, e.g., of revenue equivalence results or dominant bidding strategies.


From the algorithmic definition of the Vickrey auction we generate a library of Scala code, which includes verified code generated from everything in the Isabelle library that these definitions depend on, such as list and set functions and arithmetical operations.  These largely operate on different datatypes than those of the Scala library.  For example, the generated code expects the bidders to be a set of natural numbers, whereas a Scala program would more naturally represent input provided by a user or read from a database as a list of integers, or it represents prices as fractions of integers, whereas decimal notation would rather be used on a user interface.  Thus, to make the generated code usable in real-world applications, one needs to provide wrapper code.  Any such wrapper or user interface code needs to be written with care, as it is not verified itself, and could therefore introduce errors into the overall application.

Our formal specifications, the proofs and the generated code are freely available as part of an Auction Theory Toolbox~\cite{AuctionTheoryToolbox}, allowing their reuse in subsequent problems, such as the verification of new, related auctions.  At the time of this writing, there are paper-like as well as algorithmic formalisations of the single-good and the combinatorial Vickrey auction, a proof of “paper$\leftrightarrow$\allowbreak algorithm” equivalence as well as a proof of Vickrey's theorem in the single-good case, and proof-of-concept Scala applications with console input/\allowbreak output that execute the generated code.  In the combinatorial case the equivalence proofs on the allocation and winner determination layers remain to be finished, and we need to implement further tie-breaking rules.  On a recent laptop all proofs are checked within seconds.

\section{Related work} \label{sec:related-work}

There are three categories of work related to that in this paper: theorem proving on auctions; verification of executable auction code; confirming that an auction design is well specified by model checking.

\subsection{Theorem proving on auctions}

The authors of this paper have led, in collaboration with further theorem proving experts, a comparative study on proving Vickrey's theorem for single-good auctions in a different systems~\cite{LangeEtAl:CompProvAuctThy13}.  Isabelle/HOL performed well in this comparison, for reasons summarised in section~\ref{sec:isabelle}, but it was neither the best system w.r.t.\ the closeness of the formula input language to mathematical textbook notation nor the comprehensibility of automatically generated proofs.  Our study is incomplete in that it did not yet evaluate the systems' capabilities of managing a large, modular toolbox of formalisations, nor dealing with more advanced auction theory, e.g.\ dynamic auctions or probabilistic notions of bidding.

\subsection{Verifying auction algorithm implementations}

That the consequences of an auction's failure can be significant has been known for some time.  The Smith Institute for industrial mathematics and system engineering\footnote{\url{http://www.smithinst.co.uk/}} has “\href{http://www.smithinst.co.uk/case-study/verifying-algorithms/}{assessed for correctness the software implementations of […] algorithms}” used in UK spectrum auctions.  The results have not been published, but this work seems to involve traditional methods of running performance tests on test cases.

\subsection{Model checking and multi-agent systems}

Third and finally, another formal verification technique has been applied to auctions: model checking is not generally a substitute for theorem proving:
\begin{quote}
  Theorem proving requires a deep knowledge of mathematical structures and techniques, is usually only partly automatable, and is, in practice, often very costly, since the task can usually only be carried out by experts.  Model checking, on the other hand, relies on a complete, but automated, inspection of the system's state space. This makes it computationally costly (in terms of time and space), but does not usually require a specialist as in the case of theorem proving. In order to reduce the computational problems, state-of-the-art model checkers employ a number of reduction techniques that make it possible to handle even large state spaces relatively efficiently.~\cite{de-fi-we-bo-12}
\end{quote}

An important restriction of model checking is that -- as it involves exhaustive search -- it must operate on a finite state space.  \Citeauthor{ta-gu-va-09}~\cite{ta-gu-va-09} use SPIN, a widely-used commercial model checker based on a linear temporal logic (LTL), to verify Vickrey auctions' strategy-proofness property that bidders cannot do better than to bid their valuations.  They implemented two forms of abstraction to reduce the search space while verifying the property of interest for arbitrary bid ranges and numbers of agents: program slicing to remove variables irrelevant w.r.t.\ the property, and discretising bid values (e.g.\ “higher than someone's valuation $v_i$”).

Model checking has often been applied when modelling auctions as systems composed of multiple interacting autonomous agents.  For example, \citeauthor{xu-ch-07}~\cite{xu-ch-07} seek to detect shill bidders in concurrent auctions of identical goods, looking for suspicious bidding behaviour, including pushing prices to a reserve price before dropping out, and bidding on the higher priced version of the good.  They also use SPIN.  As detecting suspicious behaviour does not prove that a bidder is a shill, the authors assign $s$-points for particular behaviours, higher values of which are more suspicious.

\Citeauthor{de-fi-we-bo-12}~\cite{de-fi-we-bo-12} develop open source 
model checking tools, using small auctions as their primary test cases.  In a simple auction in which between three and five bidders submit sealed bids to an auctioneer, who awards a good to the highest bidder, they formally verify that the bidder with the highest bid will eventually believe that it has won the auction~\cite{we-de-fi-09}.  With even just four bidders, each of whom is restricted to making one of three possible bids, this property took just over an hour to verify.

\Citeauthor{ar-es-no-ra-si-05}~\cite{ar-es-no-ra-si-05} studied “interactions that involve autonomous, independent entities that are willing to conform to a common, explicit set of interaction conventions”.  Their leading example is a traditional open outcry auction house.  Their EIDE toolkit (electronic institution develop environment) supports both static and dynamic verification.  Static verification includes liveness checks to ensure that agents won't be blocked, that each scene (concretely: bidding round) is reachable, and that the final scene is reachable from any other, as well as correctness of the protocol (concretely: bid language).  The dynamic verification consists of simulations.

Finally, \citeauthor{mi-pa-pf-10}~\cite{mi-pa-pf-10} have developed an environment for simulating multi-agent economic environments, aiming at providing “a platform for prototyping, testing and evaluating agent-mediated markets”, and implemented the single-good Vickrey auction as an example.  As the environment is implemented in the logical programming language Prolog, which others have used for exhaustive search, we could extend it to realise a model checking like approach.  We seem to need to
\begin{enumerate}
  \item provide the missing implementations of the \texttt{get\_second\_price} and \texttt{get\_first\_bidder} functions.
  \item for any given set of bidders, sort the bid vector, $\bm{b}$, so that $b_i \ge b_{i+1}$.
  \item generate the set, $C$, of all possible bid cases, an $\left( n-1 \right)$-digit bit vector in which a “1” in the $i^{\text{th}}$ position represents $b_i > b_{i+1}$ and a “0” indicates $b_i = b_{i+1}$.
  \item for each $\bm{c} \in C$, ensure that \texttt{sealed\_auction} returns a unique outcome.
\end{enumerate}
This seems straightforward for fixed, finite $n$.  As a “proof”, it rests on our belief that $C$ correctly specifies the set of cases.  Furthermore, the approach can be modified in the special case of a second-price auction to handle arbitrary $n$ by noting that the outcome is fully determined by $b_1$ and $b_2$. 

\section{Conclusions} \label{se:concle}

We have presented a step towards applying mechanised reasoning to the design and practice of auctions.  Specifically, we confirmed that a combinatorial Vickrey auction is a function from its input (bids) to its outcome (allocation and transfers), and is therefore soundly specified.  We have also generated verified executable code that faithfully implements the combinatorial Vickrey auction.    Subsequent steps can be taken in at least two directions.  First, the present analysis can be extended to more efficient algorithmic implementations of the winner determination problem, to more sophisticated auctions, including dynamic combinatorial auctions or hybrid auctions, and -- finally -- to auction designs whose soundness remains an open question.  Second, the usability of the present method could be improved, allowing an auction designer to specify an auction in a \emph{natural language}, or to apply \emph{push button} methods for verifying its properties (rather than having to manually derive proofs), or to produce \emph{human-readable} proof that all valid input cases are satisfactorily accounted for or -- failing that -- a human-readable counter-example.

We conclude on a modest note.  First, the verification methodology presented here is in its infancy, applied to known auctions rather than to novel ones.  Thus, their utility remains to be proven in auction design.  Second, we expect that they will have only a second order effect on auction design and implementation.  The highest profile auction failures often occur at the interface with the economic environment in which they run: competitors are allowed to threaten each other, or publicly seek collusive outcomes; setting a second licence's reserve price at the first's sale price allows the winner of the first to ensure that the second is only affordable to a monopolist~\cite{kle-04}.  Preventing these lies beyond the scope of the tools presented here.  These caveats notwithstanding, the importance of auctions is such that even the second order benefits of ensuring their sound specification and correct implementation are substantial -- calling to mind Keynes' hope that economists might be seen as ``humble, competent people on a level with dentists''.

\printbibliography

\end{document}

